\documentclass{article}
\usepackage{amssymb}
\usepackage{amsmath}

\input{tcilatex}

\begin{document}

\begin{center}
\textbf{NEW\ FEATURES\ OF EXTENDED\ WORMHOLE SOLUTIONS IN THE SCALAR FIELD
GRAVITY THEORIES}

\textbf{\bigskip }

Kamal K. Nandi$^{1,2,a}$, Ilnur Nigmatzyanov$^{2,b}$, Ramil Izmailov$^{2,c}$
and Nail G. Migranov$^{2,d}$

$^{1}$\textit{Department of Mathematics, University of North Bengal,
Siliguri (W.B.) 734 013, India}

$^{2}$\textit{Joint Research Laboratory, Bashkir State Pedagogical
University, 3-A, October Revolution Str., Ufa 450000, Russia}

\textit{\bigskip }

$^{a}$Email: kamalnandi1952@yahoo.co.in

$^{b}$Email: ilnur.nigmat@gmail.com

$^{c}$Email:ramil.eijk@gmail.com

$^{d}$Email: ufangm@yahoo.co.uk

\bigskip

\textbf{Abstract}
\end{center}

\textit{The present paper reports interesting new features that wormhole
solutions in the scalar field gravity theory have. To demonstrate these, we
obtain, by using a slightly modified form of the Matos-N\'{u}\~{n}ez
algorithm, an extended class of asymptotically flat wormhole solutions
belonging to Einstein minimally coupled scalar field theory. Generally,
solutions in these theories do not represent traversable wormholes due to
the occurrence of curvature singularities. However, the Ellis I solution of
the Einstein minimally coupled theory, when Wick rotated, yields Ellis class
III solution, the latter representing a singularity-free traversable
wormhole. We see that Ellis I and III are not essentially independent
solutions. The Wick rotated seed solutions, extended by the algorithm,
contain two new parameters }$a$ and $\delta $.\textit{\ The effect of the
parameter }$a$\textit{\ on the geodesic motion of test particles reveals
some remarkable features. By arguing for Sagnac effect in the extended Wick
rotated solution, we find that the parameter }$a$ \ \textit{can indeed be
interpreted as a rotation parameter of the wormhole}. \textit{The analyses
reported here have wider applicability in that they can very well be adopted
in other theories, including in the string theory. }

\begin{center}
\bigskip

\textbf{I. Introduction}
\end{center}

Recently, there is a revival of interest in the scalar field gravity
theories including the Brans-Dicke Theory due principally to the following
reasons: Such theories occur naturally in the low energy limit of the
effective string theory in four dimensions or the Kaluza-Klein theory. It is
found to be consistent not only with the weak field solar system tests but
also with the recent cosmological observations. Moreover, the theory
accommodates Mach's principle. (It is known that Einstein's General
Relativity can not accommodate Mach's principle satisfactorily). All these
information are well known.

A less well known yet an important arena where Brans-Dicke Theory has found
immense applications is the field of wormhole physics, a field recently
re-activated by the seminal work of Morris, Thorne and Yurtsever [1].
Conceptual predecessors of modern day's wormholes could be traced to the
geometry of Flamm paraboloid, Wheeler's concept of \textquotedblleft charge
without charge\textquotedblright , Klein bottle or the Einstein-Rosen bridge
model of a charged particle [2]. Wormholes are topological handles that
connect two distant regions of space. These objects are invoked in the
investigations of problems ranging from local to cosmological scales, not to
mention the possibility of using these objects as a means of interstellar
travel [1]. Wormholes require for their construction what is called
\textquotedblleft exotic matter\textquotedblright\ - matter that violates
some or all of the known energy conditions, the weakest being the averaged
null energy condition. Such matters are known to arise in quantum effects
(Casimir effect, for example). However, the strongest theoretical
justification for the existence of exotic matter comes from the notion of
dark energy or phantom energy that are necessary to explain the present
acceleration of the Universe. Some classical fields can be conceived to play
the role of exotic matter. They are known to occur, for instance, in the $%
R+R^{2}$ theories [3], Visser's thin shell geometries [4] and, of course, in
scalar-tensor theories [5] of which Brans-Dicke theory is a prototype. There
are several other situations where the energy conditions could be violated
[6].

Brans-Dicke theory describes gravitation through a metric tensor ( $g_{\mu
\nu }$) and a massless scalar field ( $\phi $). The Brans-Dicke action for
the coupling parameter $\omega =-1$ can be obtained in the Jordan frame from
the vacuum linear string theory in the low energy limit. The action can be
conformally rescaled into what is known as the Einstein frame action in
which the scalar field couples minimally to gravity. The last is referred to
as the Einstein minimally coupled scalar field theory. Several static
(mostly nontraversable) wormhole solutions in Einstein minimally coupled
scalar field theory and Brans-Dicke theory have been widely investigated in
the literature [7]. However, to our knowledge, exact rotating wormhole
solutions are relatively scarce except a recent one in Einstein minimally
coupled scalar field theory discussed by Matos and N\'{u}\~{n}ez [8]. In
this context, we recall the well known fact that the formal independent
solutions of Brans-Dicke theory are \textit{not} unique. (Of course, the
black hole solution \textit{is} unique for which the Brans-Dicke or minimal
scalar field is trivial in virtue of the so called \textquotedblleft no
scalar hair" theorem.) Four classes of static Brans-Dicke theory solutions
were derived by Brans [9] himself way back in 1962, and the corresponding
four classes of Einstein minimally coupled field theory solutions are also
known [10]. But recently it has been shown that only two of the four classes
of Brans' solutions are independent [11]; the other two can be derived from
them. However, although all the original four classes of Brans' or Einstein
minimally coupled solutions are important in their own right, we shall here
consider, for illustrative purposes, only one of them (Ellis I) as seed
solution. The same procedure can be easily adopted in other three classes.

The general motivation in the present paper is the following: To properly
frame an algorithm for generating singularity-free asymptotically flat
rotating wormhole solutions from the Ellis seed solutions and to investigate
the role of new parameters in the extended solutions. The analyses also
answer a certain long standing query on the wormhole solutions in the
Brans-Dicke theory.

In this article, using a slightly modified algorithm of Matos and N\'{u}\~{n}%
ez [8], we shall provide a method for generating wormhole solutions from the
known static seed solutions belonging to Einstein minimally coupled scalar
field theory. The solutions can then be transferred to those of Brans-Dicke
theory via inverse Dicke transformations. For illustration of the method,
only Ellis I seed solution is considered here, others are left out because
they can be dealt with similarly. The Brans-Dicke solutions can be further
rephrased as solutions of the vacuum 4-dimensional low energy string theory
( $\omega =-1$) and the section \textbf{II }shows how to do that. \ In
sections \textbf{II-V}, we shall analyze and compare the behavior Ellis III
and the Wick rotated Ellis I solution pointing out certain interesting
differences in these geometries. In section \textbf{VI}, the study of the
geodesic motion in the extended Wick rotated Ellis I solution reveals the
meanings of the Matos-N\'{u}\~{n}ez parameter $a$ and section \textbf{VII}
shows, via consideration of the Sagnac effect, that $a$ can indeed be
accepted as a rotation parameter. Finally, in section \textbf{VIII}, we
shall summarize the results. Throughout the article, we take the signature ($%
-,+,+,+$) and units such that $8\pi G=c=1$, unless restored specifically.
Greek indices run from $0$ to $3$ while Roman indices run from $1$ to $3$.

\begin{center}
\textbf{II. The action, ansatz and the algorithm}
\end{center}

\ Let us start from the 4-dimensional, low energy effective action of
heterotic string theory compactified on a 6-torus. The tree level string
action, keeping only linear terms in the string tension $\alpha ^{\prime }$
and in the curvature $\widetilde{R}$, takes the following form in the matter
free region ( $S_{matter}=0$):%
\begin{equation}
S_{string}=\frac{1}{\alpha ^{\prime }}\int d^{4}x\sqrt{-\widetilde{g}}e^{-2%
\widetilde{\Phi }}\left[ \widetilde{R}+4\widetilde{g}^{\mu \nu }\widetilde{%
\Phi }_{,\mu }\widetilde{\Phi }_{,\nu }\right] ,
\end{equation}%
where $\widetilde{g}^{\mu \nu }$ is the string metric and $\widetilde{\Phi }$
is the dilaton field. Note that the zero values of other matter fields do
not impose any additional constraints either on the metric or on the
dilation [12]. Under the substitution $e^{-2\widetilde{\Phi }}$ $=\phi $,
the above action reduces to the BD action%
\begin{equation}
S_{BD}=\int d^{4}x\sqrt{-\widetilde{g}}\left[ \phi \widetilde{R}+\frac{1}{%
\phi }\widetilde{g}^{\mu \nu }\phi _{,\mu }\phi _{,\nu }\right] ,
\end{equation}%
in which the BD coupling parameter $\omega =-1$. This particular value is
actually model independent and it actually arises due to the target space
duality. It should be noted that the BD action has a conformal invariance
characterized by a constant gauge parameter $\xi $ [13]. Arbitrary values of
$\xi $ can actually lead to a shift from the value $\omega =-1$, but we fix
this ambiguity by choosing $\xi =0$. Under a further substitution%
\begin{eqnarray}
g_{\mu \nu } &=&\phi \widetilde{g}_{\mu \nu }  \notag \\
d\varphi &=&\sqrt{\frac{2\omega +3}{2\alpha }}\frac{d\phi }{\phi };\alpha
\neq 0;\omega \neq \frac{3}{2},
\end{eqnarray}%
in which we have introduced, on purpose, a constant parameter $\alpha $ that
can have any sign. Then the action (2) goes into the form of EMS action%
\begin{equation}
S_{EMS}=\int d^{4}x\sqrt{-g}\left[ R+\alpha g^{\mu \nu }\varphi _{,\mu
}\varphi _{,\nu }\right] .
\end{equation}%
The EMS field equations are given by%
\begin{eqnarray}
R_{\mu \nu } &=&-\alpha \varphi _{,\mu }\varphi _{,\nu } \\
\varphi _{;\mu }^{:\mu } &=&0
\end{eqnarray}%
We shall choose $\alpha =+1$, $\varphi =\varphi (l)$ in what follows. The
negative sign on the right hand side of Eq.(5) implies that the source
stresses violate some energy conditions. The ansatz we take is the following:%
\begin{equation}
ds^{2}=-f(l)\left( dt+a\cos \theta d\psi \right) ^{2}+f^{-1}(l)\left[
dl^{2}+(l^{2}+l_{0}^{2})\left( d\theta ^{2}+\sin ^{2}\theta d\psi
^{2}\right) \right] ,
\end{equation}%
where $l_{0}$ is an arbitrary constant, the constant $a$ has been
interpreted in [8] as a rotational parameter of the wormhole. We call it the
Matos-N\'{u}\~{n}ez parameter. The ansatz in (7) is actually a subclass of
the more general class of stationary metrics given by [14,15]:%
\begin{equation}
ds^{2}=-f\left( dt-\omega _{i}dx^{i}\right) ^{2}+f^{-1}h_{ij}dx^{i}dx^{j}
\end{equation}%
where the metric function $f$, the vector potential $\omega _{i}$ and the
reduced metric $h_{ij}$ depend only on space coordinates $x^{i}$. We shall
see below that the parameter $a$ can be so adjusted as to make a symmetric,
traversable wormhole out of an asymmetric, nontraversable one. Note also
that the replacement of $a\rightarrow -a$ does not alter the field equations.

The function $f(l)$ of the ansatz (7) is then a solution of the field
equations (5) and (6) if it satisfies the following:%
\begin{eqnarray}
\left[ \left( l^{2}+l_{0}^{2}\right) \frac{f^{^{\prime }}}{f}\right]
^{\prime }+\frac{a^{2}f^{2}}{l^{2}+l_{0}^{2}} &=&0, \\
\left( \frac{f^{^{\prime }}}{f}\right) ^{2}+\frac{4l_{0}^{2}+a^{2}f^{2}}{%
\left( l^{2}+l_{0}^{2}\right) ^{2}}-2\varphi ^{\prime 2} &=&0,
\end{eqnarray}%
where the prime denotes differentiation with respect to $l$.

\textbf{Algorithm:}

Let $f_{0}\equiv f_{0}(l;p,q;a=0)$ and $\varphi _{0}=\varphi _{0}(l;p,q;a=0)$
be a known seed solution set of the static configuration in which $p,$ $q$
are arbitrary constants in the solution interpreted as the mass and scalar
charge of the configuration. Then the new generated (or extended) solution
set ($f,\varphi $) is
\begin{equation}
f(l;p,q;a)=\frac{2npq\delta f_{0}^{-1}}{a^{2}+n\delta ^{2}f_{0}^{-2}}%
,\varphi (l;p,q;a)=\varphi _{0}
\end{equation}%
where $n$ is a natural number and the paremeters $p$, $q$ are specific to a
given seed solution set ($f_{0},\varphi _{0}$) while $\delta $ is a free
parameter allowed by the generated solution in the sense that it cancels out
of the nonlinear field equations. The scalar field $\varphi _{0}$ is
remarkably given by the \textit{same} static solution of the massless
Klein-Gordon equation $\varphi _{;\mu }^{:\mu }=0$. The seed solution ( $a=0$%
) following from Eqs.(9) and (10) gives $\delta =2pq$. For the generated
solution ($a\neq 0$), the value of $\delta $ may be fixed either by the
condition of asymptotic flatness or via the matching conditions at specified
boundaries. Eq.(11) is the algorithm we propose. This is similar to, but not
quite the same as, the Matos-N\'{u}\~{n}ez [8] algorithm. The difference is
that they defined the free parameter as $\delta =\sqrt{D}$ . The difficulty
in this case is that, for our seed solution set ($f_{0},\varphi _{0}$)
below, the field Eqs.(10) and (11) identically fix $\delta ^{2}=D=0$ giving $%
f=0$ which is obviously meaningless. The other difference is that we have
introduced a real number $n$ that now designates each seed solution $f_{0}$
and likewise the corresponding new solution $f$. With the known parameters $%
n $, $p$ and $q$ plugged into the right side of Eq.(11), the new solution
set ($f,\varphi $) identically satisfies Eqs. (9) and (10). One also sees
that the algorithm can be applied with the set ($f,\varphi $) as the new
seed solution and the process can be indefinitely iterated to generate
\textit{any} number of new solutions. This is a notable generality of the
algorithm.

\begin{center}
\textbf{III. Ellis I solution and its geometry }
\end{center}

The study of the solutions of the Einstein minimally coupled scalar field
system has a long history. Static spherically symmetric solutions have been
independently discovered in different forms by many authors and their
properties are well known [16,17]. We start from the following form of Class
I solution, due to Buchdahl [18], of Einstein minimally coupled theory :%
\begin{eqnarray}
ds^{2} &=&-\left( \frac{1-\frac{m}{2r}}{1+\frac{m}{2r}}\right) ^{2\beta
}dt^{2}+\left( 1-\frac{m}{2r}\right) ^{2(1-\beta )}\left( 1+\frac{m}{2r}%
\right) ^{2(1+\beta )}\times  \notag \\
&&[dr^{2}+r^{2}d\theta ^{2}+r^{2}\sin ^{2}\theta d\psi ^{2}]
\end{eqnarray}%
\begin{equation}
\varphi (r)=\sqrt{\frac{2(\beta ^{2}-1)}{\alpha }}\ln \left[ \frac{1-\frac{m%
}{2r}}{1+\frac{m}{2r}}\right] ,
\end{equation}%
where $m$ and $\beta $ are two arbitrary constants. The same solution, in
harmonic coordinates, has been obtained and analyzed also by Bronnikov [19].

The metric (13) can be expanded to give%
\begin{eqnarray}
ds^{2} &=&-\left[ 1-\frac{2M}{r}+\frac{2M^{2}}{r^{2}}+O\left( \frac{1}{r^{3}}%
\right) \right] dt^{2}+\left[ 1+\frac{2M}{r}+O\left( \frac{1}{r^{2}}\right) %
\right] \times  \notag \\
&&[dr^{2}+r^{2}d\theta ^{2}+r^{2}\sin ^{2}\theta d\psi ^{2}],
\end{eqnarray}%
from which one can read off the Keplerian mass%
\begin{equation}
M=m\beta
\end{equation}%
The solution (12) describes all weak field solar system tests because it
exactly reproduces the PPN parameters. For $\beta =1$, it reduces to the
Schwarzschild black hole solution in isotropic coordinates. For $\alpha =+1$
and $\beta >1$, it represents a naked singularity. The metric is invariant
in form under inversion of the radial coordinate $r\rightarrow \frac{m^{2}}{%
4r}$ and we have two asymptotically flat regions (at $r=0$ and $r=\infty $),
the minimum area radius (throat) occurring at $r_{0}=\frac{m}{2}\left[ \beta
+\sqrt{\beta ^{2}-1}\right] $. Thus, real throat is guaranteed by the
condition $\beta ^{2}>1$ which we might call here the wormhole condition.
However, despite these facts, because of the occurrence of naked singularity
at $r=m/2$, it is not traversable and so Visser [4] called it a
\textquotedblleft diseased" wormhole. For the choice $\alpha =+1$, the
quantity $\sqrt{2(\beta ^{2}-1)}$ is real such that there is a real scalar
charge $\sigma $ from Eq.(14) given by%
\begin{equation}
\varphi =\frac{\sigma }{r}=-\frac{2m}{r}\sqrt{\frac{\beta ^{2}-1}{2}}
\end{equation}%
But, in this case, we have violated almost all energy conditions importing
\textit{by hand }a negative sign before the kinetic term in Eq.(5).
Alternatively, we could have chosen $\alpha =-1$ in Eq.(13), which would
then give an imaginary charge $\sigma $. In both cases, however, we end up
with the same equation $R_{\mu \nu }=-\varphi _{,\mu }\varphi _{,\nu }$.
There is absolutely no problem in accommodating an imaginary scalar charge
in any configuration violating energy conditions [17,18].

The Ricci scalar $R$ for the solution (12)\ is%
\begin{equation}
R=\frac{2m^{2}r^{4}(1-\beta ^{2})}{(r-m/2)^{2(2-\beta )}(r+m/2)^{2(2+\beta )}%
}
\end{equation}%
which diverges at $r=m/2$ showing a curvature singularity there. For $\beta
\geq 2$, the divergence in the Ricci scalar is removed, but then the metric
becomes singular. However, metric singularity is often removable when one
redefinines it in better coordinates and parameters.

Using the coordinate transformation $l=r+\frac{m^{2}}{4r}$, the solution
(12) and (13) can be expressed as%
\begin{eqnarray}
ds^{2} &=&-f_{0}(l)dt^{2}+\frac{1}{f_{0}(l)}\left[ dl^{2}+(l^{2}-m^{2})%
\left( d\theta ^{2}+\sin ^{2}\theta d\psi ^{2}\right) \right] ,  \notag \\
f_{0}(l) &=&\left( \frac{l-m}{l+m}\right) ^{\beta }, \\
\varphi _{0}(l) &=&\sqrt{\frac{\beta ^{2}-1}{2}}\ln \left[ \frac{l-m}{l+m}%
\right] .
\end{eqnarray}%
In this form, it is exactly the Ellis I solution that has been discussed
also by Bronnikov and Shikin [20]. Eqs.(18), (19) identically satisfy the
field Eqs.(9) and (10) for $a=0$. This is our seed solution set but we still
need to suitably redefine it because of the apprearance of the naked
singularity at $l=m$ The throat $l_{0}$ appears at $l_{0}=r_{0}+\frac{m^{2}}{%
4r_{0}}=m\beta >m$ corresponding to $r=r_{0}$. Thus the minimum surface area
now has a value $4\pi m^{2}\beta ^{2}$. For this solution, $p=m$, $q=\beta $%
, and it turns out that the seed solutions (18), (19) correspond to $n=4$ so
that the generated EMS solution that identically satisfies the field
equations (9) and (10) for $a\neq 0$ are:%
\begin{equation}
f(l;m,\beta ,a)=\frac{8m\beta \delta f_{0}^{-1}}{a^{2}+4\delta ^{2}f_{0}^{-2}%
};\varphi (l)=\sqrt{\frac{\beta ^{2}-1}{2}}\ln \left[ \frac{l-m}{l+m}\right]
.
\end{equation}%
To achieve asymptotic flatness at both sides, that is, $f(l)\rightarrow 1$
as $l\rightarrow \pm \infty $, we note that $f_{0}(l)\rightarrow 1$ as $%
l\rightarrow \pm \infty $ . Therefore, we must fix%
\begin{equation}
\delta =\frac{2M\pm \sqrt{4M^{2}-a^{2}}}{2}.
\end{equation}%
In the above, we should retain only the positive sign before the square
root. The reason is that, for $a=0$, the negative sign gives $\delta =0$
implying $f=0$ which is meaningless. On the other hand, the positive root
gives $\delta =2M$ and $f=f_{0}$, as desired. Note that $\beta =1$ does not
lead to Kerr black hole solution from the generated metric. Therefore, the
latter does not represent a rotating black hole but might represent the
spacetime of a rotating wormhole [8].

Let us now examine wormhole geometries in the static and generated solutions
in the EMS.

\textit{(a) Static seed case (}$a=0$\textit{):}

The first observation is that the metric functions in Eq.(18) diverge at the
singularity $l=\pm m$ as does the Ricci scalar. The next observation relates
to the behavior of the area radius. It exhibits certain peculiar properties
for the metric in (18) for $\beta >1$. For the segments $l\geq m$ $\ $and $%
l\leq -m$, we have the area radius $\rho _{0}^{I}(l)=\sqrt{%
f_{0}^{-1}(l^{2}-m^{2})}$. Then, the area $4\pi \rho _{0}^{2}(l)$ decreases
from $+\infty $ at one asymptotic flat end to a minimum value $\rho _{0\min
}=\rho _{0}(l_{0})$ at the throat $l=l_{0}=m\beta $, and then becomes
asymptotically large, but \textit{not} flat, at a radial point $l=m$. In the
remaining segment, we have $\left\vert l\right\vert <m$, and the area now
has to be redefined as $\rho _{0}^{II}(l)=\sqrt{f_{0}^{-1}\left\vert
m^{2}-l^{2}\right\vert }$. It then decreases to zero at $l=-m$ (another
throat, zero radius!) and then opens asymptotically out to $-\infty $ at the
other asymptotic flat end. Though in the $r-$coordinate version, the metric
is inversion symmetric, there are now two asymptotically flat, isometric
universes with their own throats and are actually \textit{disconnected} by
the naked singularity at $l=m$. They are also \textit{asymmetric} around $%
l=0 $ due to the fact that the throat radii in the two universes are
different.\

\textit{(b) Generated case (}$a\neq 0$\textit{):\ }

Here again, the metric functions given by Eq.(20) diverge at $l=\pm m$. The
throat of the rotating wormhole can be found from the roots of the equation
for $l:$%
\begin{equation}
a^{2}[l(f_{0}^{2}-1)+m\beta ((f_{0}^{2}+1)]+4m\beta (l-m\beta )(2m\beta +%
\sqrt{4m^{2}\beta ^{2}-a^{2}})=0
\end{equation}%
They can be computed only numerically for given values of the parameters $m$
and $\beta $. However, the area radius $\rho (l)=\sqrt{f^{-1}(l^{2}-m^{2})}$
for the solution $f$ shows that, for $a\neq 0$, the area jumps to infinity
for $\beta >1$ at $l=\pm m$ but flares out asymptotically to $\pm \infty $
on both sides. One now has three disconnected universes, that is, a
one-sided asymptotically flat universe, a both-sided non-flat but
asymptotically large \textquotedblleft sandwich universe" and another
one-sided asymptotically flat universe. For the extreme case $a=2m\beta $,
the picture is the same but the behavior of $\rho (l)$ is symmetric around $%
l=0$. Thus, the wormhole is not traversable, be it seed solution (18) or
extended solution (20). A natural question arises if they be made
traversable by removing the singularity manifested in the infinite jump in
the area at $l=\pm m$ and in the curvature. \ We shall consider this
question now.

\begin{center}
\textbf{IV. Ellis III solution via Wick rotation}\
\end{center}

One procedure to remove the aforementioned singularities is to analytically
continue the Ellis I solutions ($f_{0}$,$\varphi _{0}$) by means of Wick
rotation of the parameters while maintaining the real numerical value of the
throat radius. In the solution set ($f_{0}$,$\varphi _{0}$), we choose
\begin{equation}
m\rightarrow -im,\beta \rightarrow i\beta
\end{equation}%
so that $l_{0}=m\beta $ is invariant in sign and magnitude.

\textit{(a) }$a=0$\textit{:}

Then the metric resulting from the seed Eq.(18) is and it is our redefined
seed solution:%
\begin{equation}
ds^{2}=-f_{0}^{\prime }(l)dt^{2}+\frac{1}{f_{0}^{\prime }(l)}\left[
dl^{2}+(l^{2}+m^{2})\left( d\theta ^{2}+\sin ^{2}\theta d\psi ^{2}\right) %
\right]
\end{equation}%
\begin{equation}
f_{0}^{\prime }(l)=\exp \left[ -2\beta \func{arccot}\left( \frac{l}{m}%
\right) \right]
\end{equation}%
\begin{equation}
\varphi _{0}^{\prime }(l)=\left[ \sqrt{2}\sqrt{1+\beta ^{2}}\right] \func{%
arccot}\left( \frac{l}{m}\right)
\end{equation}%
This is no new solution but is just the Ellis III solution [19] which can be
obtained in the original form by using the relation [20]%
\begin{eqnarray}
\func{arccot}(x)+\arctan (x) &=&+\frac{\pi }{2};x>0 \\
&=&-\frac{\pi }{2};x<0
\end{eqnarray}%
and the function on the left shows a finite jump (of magnitude $\pi $) at $%
x=0$. Thus, we get from Eqs.(25) and (26) two branches, the $+ve$ sign
corresponds to the side $l>0$ and the $-ve$ sign to $l<0$ [22]:

\begin{equation}
f_{0\pm }^{Ellis}(l)=\exp [-2\beta \{\pm \frac{\pi }{2}-\arctan (\frac{l}{m}%
)\}]
\end{equation}%
\begin{equation}
\varphi _{0\pm }^{Ellis}(l)=\left[ \sqrt{2}\sqrt{1+\beta ^{2}}\right] \left(
\pm \frac{\pi }{2}-\arctan (\frac{l}{m})\right)
\end{equation}%
We might study the solutions (30) and (31) \textit{per se}, or equivalently,
study the two restricted branches taken together, while allowing for a
discontinuity at the origin $l=0$. Otherwise, we might disregard (30), (31)
and treat \textit{each} of the $\pm $ set in Eqs.(34), (35) as independently
derived exact solution valid in the unrestricted range of $l$ with no
discontinuity at $l=0$. The two alternatives do not appear quite the same.
In fact, each of the individual branch represents a geodesically complete,
asymptotically flat traversable wormhole (termed as "drainholes" by Ellis)
having different masses, one positive and the other negative, on two sides
respectively. The known Ellis III solution is the $+ve$ branch which is
continuous over the entire interval $l\in (-\infty ,+\infty )$. The $-ve$
branch is also equally good. What we have shown here is that the Ellis
solutions I and III are \textit{not} independent solutions of the Einstein
minimally coupled scalar field theory as one can be obtained from the other.

It is of interest to compare the behaviors of the Ellis III solutions (34)
with the Wick rotated Ellis I solutions (30):\ (i) The Ellis III metric
function $f_{0+}^{Ellis}(l)\rightarrow 1$ as $l\rightarrow +\infty $ but $%
f_{0+}^{Ellis}(l)\rightarrow e^{-2\pi \beta }$ as $l\rightarrow -\infty $.
These two limits correspond to a Schwarzschild mass $M$ at one mouth and $%
-Me^{\pi \beta }$ at the other. There is no discontinuity at the origin
because $f_{0+}^{Ellis}(l)\rightarrow e^{-\pi \beta }$ as $l\rightarrow \pm
0 $. In the solution (30), on the other hand, there is a discontinuity at
the origin because $f_{0}(l)\rightarrow e^{\pm \pi \beta }$ as $l\rightarrow
\pm 0$ while there is no asymptotic mass jump since $f_{0}(l)\rightarrow 1$
as $l\rightarrow \pm \infty $. The curvature scalars for both (30) and (34)
are formally the same and given by%
\begin{eqnarray}
R_{0}^{\prime } &=&-\frac{2m^{2}(1+\beta ^{2})}{(l^{2}+m^{2})^{2}}\exp \left[
-2\beta \func{arccot}\left( \frac{l}{m}\right) \right] \\
R_{0+}^{Ellis} &=&-\frac{2m^{2}(1+\beta ^{2})}{(l^{2}+m^{2})^{2}}\exp \left[
-2\beta \{\frac{\pi }{2}-\arctan \left( \frac{l}{m}\right) \}\right]
\end{eqnarray}%
which go to zero as $l\rightarrow \pm \infty $. That means, the spacetime is
flat on two sides for both the solutions. Next, we verify what happens to
these scalars at the singular coordinate radius ($r=m/2$) that has now been
shifted to the origin $l=r-\frac{m^{2}}{4r}=0$. (ii) The Ellis curvature
scalar $R_{0+}^{Ellis}\rightarrow -\frac{2(1+\beta ^{2})}{m^{2}}e^{-\pi
\beta }$ as $l\rightarrow \pm 0$ whereas the curvature scalar $R_{0}^{\prime
}$ exhibits a finite jump from $-\frac{2(1+\beta ^{2})}{m^{2}}e^{-\pi \beta
} $ to $-\frac{2(1+\beta ^{2})}{m^{2}}e^{+\pi \beta }$ as $l\rightarrow \pm
0 $. (iii) The area radius $\rho _{0+}^{Ellis}(l)=\sqrt{%
f_{0+}^{-1(Ellis)}(l^{2}+m^{2})}\rightarrow m\sqrt{e^{\pi \beta }}$ as $%
l\rightarrow \pm 0$ whereas $\rho _{0}^{\prime }(l)=\sqrt{f_{0}^{\prime
-1}(l^{2}+m^{2})}$ shows a finite jump from $m\sqrt{e^{\pi \beta }}$ to $m%
\sqrt{e^{-\pi \beta }}$ as $l\rightarrow \pm 0$. These show that while Ellis
wormhole (34) is traversable, but jumps at the origin in the Wick rotated
solution (30) prevent traversability. The behaviors of (34) and (30) are
different at the origin except that, for both the solutions, the throat
appears at the same radius $l_{0}=M=m\beta $. Similar considerations apply
for the $-ve$ branch.

Ellis III wormholes (34) can be straightforwardly extended to the rotating
form via the algotithm (11) and they would be likewise traversable, as has
been shown by Matos and N\'{u}\~{n}ez [8]. Thus we do not deal with it here.
Henceforth, we would rather concentrate on the Wick rotated Ellis I solution
and ask: Can we somehow remove the discontinuities in (30)? That is exactly
where the new paramter $a$ comes in. Let us see what happens when $a\neq 0$.

\textit{(b) }$a\neq 0:$

The minimum area radius of the extended solution is obtained from the
equation $\frac{d\mu ^{\prime }}{dl}=0$ where $\rho ^{\prime }(l)=\sqrt{%
f^{\prime -1}(l^{2}+m^{2})}$ is the area radius. From numerical study of the
resulting equation, we find that the minimum area occurs at $l<m\beta $ and
it decreases with the increase of $a$ for fixed values of $m$ and $\beta $.
We also notice that the finite jump persists in the area radius of the
extended solution $f^{\prime }(l;m,\beta ,a)$, $\delta $ being still given
by Eq.(21). Surprisingly however, when $a=2m\beta $, the area function $\rho
^{\prime }(l)$ decreases from $+\infty $ to the minimum value at the throat,
then increases to a finite value at $l=0$, undergoes no jump at $l=0$ but
passes continuously, though not with $C^{2}$ smoothness, across $l=0$, on to
$-\infty $. The Ricci scalar $R^{\prime }$ for $f^{\prime }(l;m,\beta ,a)$
is given by
\begin{equation}
R^{\prime }=\frac{8\beta \left( 1+\beta ^{2}\right) (2m\beta +\sqrt{%
4m^{2}\beta ^{2}-a^{2}})\exp [2\beta \func{arccot}(\frac{l}{m})]}{m\left( 1+%
\frac{l^{2}}{m^{2}}\right) ^{2}\left[ a^{2}+\left( 2m\beta +\sqrt{%
4m^{2}\beta ^{2}-a^{2}}\right) ^{2}\exp [4\beta \func{arccot}(\frac{l}{m})]%
\right] }
\end{equation}%
and it approaches the value $\frac{4(1+\beta ^{2})e^{\pi \beta }}{%
m^{2}(1+e^{2\pi \beta })}$ as $l\rightarrow \pm 0$, that is, no jump in it.

The area behavior shows that, for the extreme case $a=2m\beta $, we do have
a traversable wormhole with a \textit{single} metric covering both the
asymptotically flat universes ($l\rightarrow \pm \infty $) connected by a
finite wedge-like protrusion in the shape function at $l=0$. This wedge
prevents $C^{2}$ continuity across $l=0$ in the area function but sews up
two exactly symmetrical asymptotically flat universes on both sides. The
numerical values of the free parameters $m$ and $\beta $ can always be
suitably controlled to make the tidal force humanly tolerable and the travel
safer.

For $a\neq 2m\beta $, such a single coordinate chart is not possible as the
area has a jump at $l=0$. However, we can artificially circumvent this
discontinuity and connect the two disjoint universes by multiple metric
choices on different segments. We can get a cue for this construction from
the static case. Consider the metric form (18) on one segment ($AB$) and the
Wick rotated metric (24) on the other side ($BC$) so that the areas match at
a radial point $l=l_{1}$. The radius $l=l_{1}$ is a root of the equation
(area from right $(AB)$ = area from left $(BC)$)
\begin{equation}
(m^{2}-l^{2})\times \left( \frac{l-m}{l+m}\right) ^{-\beta
}=(m^{2}+l^{2})\times Exp[2\beta \func{arccot}(\frac{l}{m})]
\end{equation}%
By numerical computation, it turns out that $0<l_{1}<m$ such that the two
otherwise disjoint universes, one represented by the branch $AB$ and the
other by $BC$, can be connected at $B(l=l_{1})$. At the joining point $B$,
there is continuity in the area function (again not $C^{2}$) and the tidal
forces can be shown to be finite throughout the generator curve $ABC$.
Exactly similar arguments hold in the rotating case. Branche $AB$ belongs to
the Wick rotated solution ($f^{\prime }$) while the sector $BC$ belongs to
the original solution ($f$). Numerical calculations show that the matching
occurs at either of the two points $B(l_{1})$ or $\allowbreak B(l_{2})$ such
that $-m<l_{1},l_{2}<m$.

Traversable wormholes can also be constructed by employing the
\textquotedblleft cut-and-paste" procedure [4]. One takes two copies of the
static wormholes and joins them at a radius $l=l_{b}>l_{0}$. The interface
between the two copies will then be described by a thin shell of exotic
matter. The shape functions on both sides will be symmetric. However, when
rotation is introduced, numerical calculation shows that the throat radius
decreases from the static value while the flaring out occurs faster. It is
of some interest to note that Cris\'{o}stomo and Olea [20,21] developed a
Hamiltonian formalism to obtain the dynamics of a massive rotating thin
shell in (2+1) dimensions. There, the matching conditions are understood as
continuity of the Hamiltonian functions for an ADM foliation of the metric.
Of course, this procedure can be trivially extended to deal with axially
symmetric solutions in (3+1) dimensions. For soliton solutions, see Ref.[22].

\begin{center}
\textbf{V. Extended Brans-Dicke I solution}
\end{center}

To obtain the rotating Brans-Dicke solution, we pursue the following steps:
Note from Eq.(3) that%
\begin{equation}
\sqrt{\frac{2\omega +3}{2}}\ln \phi =\varphi =\ln \left[ \frac{1-\frac{m}{2r}%
}{1+\frac{m}{2r}}\right] ^{\sqrt{2(\beta ^{2}-1)}}\Rightarrow \phi =\left[
\frac{1-\frac{m}{2r}}{1+\frac{m}{2r}}\right] ^{\sqrt{4(\beta
^{2}-1)/(2\omega +3)}}.
\end{equation}%
Now using the constraint from the Brans-Dicke field equations [9], viz.,%
\begin{equation}
4(\beta ^{2}-1)=-(2\omega +3)\frac{C^{2}}{\lambda ^{2}},
\end{equation}%
where $C,\lambda $ are two new arbitrary constants and $\omega $ is the
coupling parameter, we get%
\begin{equation}
\phi =\left[ \frac{1-\frac{m}{2r}}{1+\frac{m}{2r}}\right] ^{\frac{C}{\lambda
}}=\left[ \frac{l-m}{l+m}\right] ^{\frac{C}{2\lambda }}.
\end{equation}%
The Eq.(34) can be rephrased in the familiar form [9]:%
\begin{equation}
\lambda ^{2}=(C+1)^{2}-C(1-\frac{\omega C}{2}).
\end{equation}%
However, the minimum area condition $\beta ^{2}>1$ requires that the right
hand side of Eq.(34) be positive. This is possible if either $\omega <-\frac{%
3}{2}$ or $\lambda $ be imaginary. Let us first consider $\omega <-\frac{3}{2%
}$ so that the exponents are real. Then, the final step consists in using
the relation $\widetilde{g}_{\mu \nu }=\phi ^{-1}g_{\mu \nu }$ together with
replacing $\beta $ in the exponents in the $g_{\mu \nu \text{ }}$ by [7]%
\begin{equation}
\beta =\frac{1}{\lambda }\left( 1+\frac{C}{2}\right) .
\end{equation}%
This means we have the Brans-Dicke rotating class I solution for $\omega <-%
\frac{3}{2}$ as follows:%
\begin{eqnarray}
ds^{2} &=&-\widetilde{f}_{1}(l)dt^{2}+\widetilde{f}_{2}(l)\left[
dl^{2}+(l^{2}-m^{2})\left( d\theta ^{2}+\sin ^{2}\theta d\psi ^{2}\right) %
\right] , \\
\widetilde{f}_{1}(l) &\equiv &\widetilde{f}_{1}(l;m,C,\lambda
,a)=f(l;m,\beta ,a)\phi ^{-1}  \notag \\
&=&\frac{8m\delta \left[ \frac{1}{2\lambda }(C+2)\right] \left[ \frac{l-m}{%
l+m}\right] ^{-\frac{1}{\lambda }\left( 1+\frac{C}{2}\right) }}{%
a^{2}+4\delta ^{2}\left[ \frac{l-m}{l+m}\right] ^{-\frac{2}{\lambda }\left(
1+\frac{C}{2}\right) }}\times \left[ \frac{l-m}{l+m}\right] ^{-\frac{C}{%
2\lambda }}, \\
\widetilde{f}_{2}(l) &\equiv &f_{2}(l;m,C,\lambda ,a)=f^{-1}(l;m,\beta
,a)\phi ^{-1}  \notag \\
&=&\frac{a^{2}+4\delta ^{2}\left[ \frac{l-m}{l+m}\right] ^{-\frac{2}{\lambda
}\left( 1+\frac{C}{2}\right) }}{8m\delta \left[ \frac{1}{2\lambda }(C+2)%
\right] \left[ \frac{l-m}{l+m}\right] ^{-\frac{1}{\lambda }\left( 1+\frac{C}{%
2}\right) }}\times \left[ \frac{l-m}{l+m}\right] ^{-\frac{C}{2\lambda }}, \\
\phi (l) &=&\left[ \frac{l-m}{l+m}\right] ^{\frac{C}{2\lambda }}.
\end{eqnarray}%
It can be verified that the Brans-Dicke field equations again yield the
expression (36). Using the relation $l=r+\frac{m^{2}}{4r}$, it can be easily
expressed in the familiar ($t,r,\theta ,\psi $) coordinates with the value
of $\delta $ given by Eq.(21) in which $\beta $ should have the value as in
Eq.(36). For instance, when $a=0$, we have $\delta =\frac{m}{\lambda }(C+2)$
and identifying $\frac{m}{2}=B$, one retrieves the static Brans-Dicke metric
in the original notation:%
\begin{eqnarray}
ds^{2} &=&-\left( \frac{1-\frac{B}{r}}{1+\frac{B}{r}}\right) ^{\frac{2}{%
\lambda }}dt^{2}+\left( 1+\frac{B}{r}\right) ^{4}\left( \frac{1-\frac{B}{r}}{%
1+\frac{B}{r}}\right) ^{\frac{2(\lambda -C-1)}{\lambda }}\times  \notag \\
&&[dr^{2}+r^{2}d\theta ^{2}+r^{2}\sin ^{2}\theta d\psi ^{2}] \\
\phi (r) &=&\left( \frac{1-\frac{B}{r}}{1+\frac{B}{r}}\right) ^{\frac{C}{%
\lambda }}.
\end{eqnarray}%
The solution (42) represents a naked singularity at $r=B$ and the condition
for the existence of a minimum area is [6,7]
\begin{equation}
(C+1)^{2}>\lambda ^{2}.
\end{equation}%
For $\beta ^{2}>1$, and $\alpha =+1$, the negative kinetic term in the field
equations (5) shows that the energy density is negative violating at least
the Weak Energy Condition. The solution (38) still does not represent a
traversable rotating wormhole in the Einstein minimally coupled theory. One
has to first make a change%
\begin{equation}
m\rightarrow im,\lambda \rightarrow -i\lambda
\end{equation}%
in (38) and (41) to get the Wick rotated counterpart. The next step is to
use the relations (32), (33) obtaining two branches as obtained in Sec.IV.
Either of the branches would then represent rotating traversable wormholes
in the Brans-Dicke theory for the range of coupling values $\omega <-\frac{3%
}{2}$. Same classes of solutions will be obtained by alternative
calculations with $\beta ^{2}>1$ and $\alpha =-1$ (imaginary scalar charge).
The above steps represent the basic scheme to be followed in other classes
of solutions in the Einsten minimally coupled or Brans-Dicke theory, which
we do not do here.

The discussion about traversability of the wormholes in the Einstein
minimally coupled theory can be transferred almost \textit{in verbatim} to
the Brans-Dicke theory, once we use the crucial relation (37) connecting the
parameters in both the theories.

\begin{center}
\textbf{VI. Geodesic motion in the extended solution}
\end{center}

We shall consider, as an illustration, the class of solutions (24) generated
from the Wick rotated seed solutions (18) of Einstein minimally coupled
theory. The resulting solution is (now dropping primes)%
\begin{eqnarray}
f(l;m,\beta ,a) &=&\frac{8m\beta \delta \exp [2\beta \func{arccot}(\frac{l}{m%
})]}{a^{2}+4\delta ^{2}\exp [4\beta \func{arccot}(\frac{l}{m})]} \\
\varphi (l) &=&\left[ \sqrt{2(1+\beta ^{2})}\right] \func{arccot}\left(
\frac{l}{m}\right)
\end{eqnarray}%
so that the metric components $g_{\nu \lambda }$ are%
\begin{eqnarray}
g_{00} &=&-f, \\
g_{11} &=&f^{-1}, \\
g_{22} &=&f^{-1}(l^{2}+l_{0}^{2}), \\
g_{33} &=&f^{-1}(l^{2}+l_{0}^{2})\sin ^{2}\theta -fa^{2}\cos ^{2}\theta , \\
g_{03} &=&g_{30}=-fa\cos \theta .
\end{eqnarray}%
The four velocity is defined by
\begin{equation}
U^{\mu }=\frac{dx^{\mu }}{dp},x^{\mu }\equiv (t,l,\theta ,\phi ),dp=m_{0}ds
\end{equation}%
in which $p$ is the new affine parameter and $m_{0}$ is the invariant rest
mass of the test particle. The geodesic equations are given by%
\begin{equation}
\frac{dU_{\mu }}{dp}-\frac{1}{2}\frac{\partial g_{\nu \lambda }}{\partial
x^{\mu }}U^{\nu }U^{\lambda }=0,g_{\nu \lambda }U^{\nu }U^{\lambda
}=m_{0}^{2}=\epsilon
\end{equation}%
Since the metric functions $g_{\nu \lambda }$ do not contain $t$ and $\phi $%
, the corresponding momenta are conserved, that is, the $\mu =0$ and $\mu =3$
equations give, respectively%
\begin{eqnarray}
U_{0} &=&-f.U^{0}-faU^{3}\cos \theta =k \\
U_{3} &=&-faU^{0}\cos \theta +U^{3}[f^{-1}(l^{2}+l_{0}^{2})\sin ^{2}\theta
-fa^{2}\cos ^{2}\theta ]=h
\end{eqnarray}%
where $k$ and $h$ are arbitrary constants. From the above, it follows that
the \textquotedblleft angular momentum\textquotedblright\ of the particle is%
\begin{eqnarray}
r^{2}(l)U^{3} &=&\frac{h-ka\cos \theta }{\sin ^{2}\theta }\equiv \xi \\
r^{2}(l) &\equiv &f^{-1}(l^{2}+l_{0}^{2}).
\end{eqnarray}%
The $\mu =2$ component of the geodesic equation gives%
\begin{equation}
\frac{d}{dp}\left( r^{2}(l)\frac{d\theta }{dp}\right) +\frac{1}{2}%
faU^{0}U^{3}\sin \theta -(U^{3})^{2}[r^{2}(l)+fa^{2}]\cos \theta \sin \theta
=0.
\end{equation}%
Instead of the $\mu =1$ equation, we take the second of Eq.(53) which gives%
\begin{equation}
\epsilon =-f[U^{0}+U^{3}a\cos \theta
]^{2}+f^{-1}(U^{1})^{2}+r^{2}(l)[(U^{2})^{2}+(U^{3})^{2}\sin ^{2}\theta ].
\end{equation}%
Looking at Eq.(58), we see that two solutions are possible. One is
\begin{equation}
U^{3}=0\Rightarrow \theta =\arccos (\frac{h}{ka})=const.\Rightarrow U^{2}=0
\end{equation}%
which means $\theta $ can assume any constant value depending on the
independent values of $h,k$ and $a$. But such motions will only be radial
since $U^{2}=0$ and $U^{3}=0$. In other words the gravitating source acts
like a \textit{radial sink}! The geodesic equation of motion\ (59) reduces
to:%
\begin{equation}
\left( \frac{dl}{dp}\right) ^{2}=\epsilon f(l;m,\beta ,a)+k^{2}
\end{equation}%
This can be rewritten very succinctly as%
\begin{equation}
\frac{d^{2}l}{dp^{2}}=\frac{\epsilon }{2}\frac{df}{dl}.
\end{equation}

The other solution of Eq.(58) is $\theta =0$, which implies that the test
particle motion is restricted to polar planes. However, we can always choose
the pole perpendicular to this plane so that the angle $\varphi $ varies in
that plane with the particle motion. Moreover, from Eq.(56), we must have $%
h=ka$ so that $r^{2}(l)U^{3}=\xi \neq 0.$Then, we end up again with the same
metric function but without the explicit appearance of $a$ as an arbitrarily
regulated free parameter. Thus, in the equation of motion, we have $%
f=f(l;m,\beta ,h/k)$ \ so that $a$, which is a parameter of the gravitating
source, is obtained from the \textit{motional characteristics} like $h$ and $%
k$ of the test particle itself. This intriguing feature is somewhat
analogous to the fact that the mass of the gravitating Sun can be determined
from the motion of a test particle (planet) around it. As a special case,
the parameter $a$ can be so chosen as to completely mask the angular
momentum $\xi $ of the test particle, that is, as $h\rightarrow ka$, there
is a possibility that $\xi \rightarrow 0$ too. Physically, it is like
choosing the parameter $a$ to coincide with the orbital $\varphi -$angular
momentum of the test particle. The equation of motion\ is again exactly the
same as Eq.(59) since $U^{2}=0$ even though $U^{3}$($\neq 0$) does not
appear explicitly. This is due to the fact that, in the last term of
Eq.(58), $(U^{3})^{2}\sin ^{2}\theta =0$ but the signature of orbiting
(non-radial) test particle is manifest in the presence of $h$:%
\begin{equation}
\left( \frac{dl}{dp}\right) ^{2}=\epsilon f(l;m,\beta ,h/k)+k^{2}
\end{equation}%
The turning points of the orbit will occur when $\epsilon f=-k^{2}$ and $%
\frac{df}{dl}\neq 0$. From these conditions, we have the turning points
occurring at%
\begin{equation}
l=l_{0}=m\cot \left( \frac{\ln x_{0}}{\beta }\right)
\end{equation}%
where%
\begin{equation}
x_{0}=\frac{-2m\beta \epsilon \pm \sqrt{4m^{2}\beta ^{2}\epsilon
^{2}-k^{4}a^{2}}}{2k^{2}\delta }.
\end{equation}%
Circular orbits will occur if $\epsilon f=-k^{2}$ and $\frac{df}{dl}=0$ and
they will be stable if $\frac{d^{2}f}{dl^{2}}<0$ and unstable if $\frac{%
d^{2}f}{dl^{2}}>0$.

\begin{center}
\textbf{VII. Sagnac effect}
\end{center}

The best way to assess the effect of a non-zero $U^{3}$ (or $d\varphi \neq 0$%
) and of the Matos-N\'{u}\~{n}ez parameter $a$ is through the Sagnac effect
[23] analyzing the orbit in the plane $\theta =0$. The effect stems from the
basic physical fact that the round trip time of light around a closed
contour, when the source is fixed on a turntable, depends on the angular
velocity, say $\Omega $, of the turntable. Using special theory of
relativity and assuming $\Omega r\ll c$, one obtains the proper time $\delta
\tau _{s}$, when the two beams meet again at the starting point as%
\begin{equation}
\delta \tau _{s}\cong \frac{4\Omega }{c^{2}}S
\end{equation}%
where $S$ ($\equiv \pi r^{2}$) is the projected area of the contour
perpendicular to the axis of rotation.

Without any loss of rigor, we take $a\rightarrow -a$ for notational
convenience although it is not mandatory. Suppose that the source/receiver
of two oppositely directed light beams is moving along a circumference $l=R=$
constant. Suitably placed mirrors reflect back to their origin both beams
after a circular trip about the central rotating wormhole. (The motion is
thus \textit{not} geodesic or force free!). Let us further assume that the
source/receiver is moving with uniform orbital angular speed $\omega _{0}$
with respect to distant stars such that the rotation angle is%
\begin{equation}
\varphi _{0}=\omega _{0}t.
\end{equation}%
Under these conditions, the metric becomes%
\begin{equation}
d\tau ^{2}=-f(R;m,\beta ,a)[1-a\omega _{0}]^{2}dt^{2}
\end{equation}%
The trajectory of a light ray is $d\tau ^{2}=0$ which gives%
\begin{equation}
\lbrack 1-a\omega ]^{2}=0
\end{equation}%
where $\omega $ is the angular speed and where we have suspended the
condition $h=ka$ for photon orbits. The roots of the above equation coincide
and is%
\begin{equation}
\omega _{1\pm }=\frac{1}{a}.
\end{equation}%
Therefore, the rotation angle for the light rays is
\begin{equation}
\varphi =\omega _{1\pm }t=\omega _{1\pm }\frac{\varphi _{0}}{\omega _{0}}.
\end{equation}%
The first intersection of the world lines of the two light rays with the
world line of the orbiting observer after emission at time $t=0$ occurs when%
\begin{equation}
\varphi _{+}=\varphi _{0}+2\pi ,\varphi _{-}=\varphi _{0}-2\pi
\end{equation}%
so that%
\begin{equation}
\frac{\varphi _{0}}{\omega _{0}}\omega _{1\pm }=\varphi _{0}\pm 2\pi
\end{equation}%
where $\pm $ refer to co-rotating and counter-rotating beams respectively.
Solving for $\varphi _{0}$, we get%
\begin{equation}
\varphi _{0\pm }=\pm \frac{2\pi \omega _{0}}{\omega _{1\pm }-\omega _{0}}.
\end{equation}%
The proper time as measured by the orbiting observer is found from Eq.(66)
by using $dt=d\varphi _{0}/\omega _{0}$ and integrating between $\varphi
_{0+}$ and $\varphi _{0-}$. The final result is the Sagnac delay given by%
\begin{eqnarray}
\left\vert \delta \tau _{s}\right\vert &=&\sqrt{f(R;m,\beta ,a)}\left[ \frac{%
1-a\omega _{0}}{\omega _{0}}\right] (\varphi _{0+}-\varphi _{0-})  \notag \\
&=&\sqrt{f(R;m,\beta ,a)}\left[ \frac{1-a\omega _{0}}{\omega _{0}}\right] %
\left[ \frac{4\pi \omega _{0}}{\frac{1}{a}-\omega _{0}}\right] .  \notag \\
&=&\sqrt{f(R;m,\beta ,a)}(4\pi a).
\end{eqnarray}%
This shows remarkably that the Sagnac delay depends only on the Matos-N\'{u}%
\~{n}ez parameter $a$. Interestingly, the result is independent of $\omega
_{0}$ meaning that it is \textit{independent }of the motional state of the
source/receiver, be it static with respect to the frame of distant stars or
moving with regard to it. When $a=0$, there is no delay because the wormhole
spacetime is then nonrotating (no turntable!). The above result supports the
conclusion of Ref.[8] from an altogether different viewpoint that $a$ can
indeed be interpreted as a rotational parameter of the wormhole.

\begin{center}
\textbf{VIII. Summary }
\end{center}

Asymptotically flat rotating solutions are rather rare in the literature, be
they wormholes or naked singularities. The algorithm (11), together with
some operations, provides a method to generate new traversable wormhole
solutions in the Einstein minimally coupled and then in the Brans-Dicke
theory. The present study opens up possibilities to explore in more detail
new solutions in other theories too. For instance, \ the string solutions
are just the Brans-Dicke solutions with $\omega =-1$. As we saw, the
asymptotically flat wormhole solutions admit two arbitrary parameters $a$
and $\delta $. The Matos-N\'{u}\~{n}ez parameter $a$ has been interpreted by
the authors (Ref [8]) as a rotation parameter. Here we have shown how the
parameters can be adjusted for obtaining traversability.

The static Ellis I wormholes do not appear traversable due to the
singularity as manifested in the behavior of the area function and
curvature. To tackle this problem, we analytically continued it via Wick \
rotation and rederived Ellis III singularity free, asymptotically flat,
traversable wormholes as one of its branches. This showed that the two
classes of solutions are not independent, contrary to the general belief.
The comparative features of the Wick rotated and Ellis III solutions are
pointed out. In the extended Wick rotated solution, numerical graphics
showed that the wormhole can be covered by a single metric with $C^{0\text{ }%
}$ continuity. That is, the jumps can be sewed up at the origin for the
extreme value of $a$ ($=2m\beta $) or can be avioded by choosing multiple
metric patches for nonextreme values of $a$, the junctions again having only
$C^{0}$ continuity. In either case, the tidal forces depend on controllable
parameters $m$ and $\beta $ so that practical traversability is assured. As
an illustrative seed solution, we considered only the Ellis I solution in
the Einstein minimally coupled scalar field theory and finally mapped the
extended solution into that of Brans-Dicke theory.

In connection with the static Brans-Dicke I solution, we recall a long
standing query by Visser and Hochberg [24] which has been the guiding
motivation in Secs. II-V: \textit{\textquotedblleft It would be interesting
to know a little bit more about what this region actually looks like, and to
develop a better understanding of the physics on the other side }[that is,
across the naked singularity] \textit{of this class of Brans-Dicke
wormholes."} \ The analyses above answer how one could achieve a both sided
asymptotically flat traversable singularity free wormhole via Wick rotation
of the Ellis I solution which is merely the conformally rescaled Brans-Dicke
I solution. \ Because of the fact that Ellis III is rederivable from Ellis I
and that the former is a traversable wormhole, we can say that Brans-Dicke I
solution is also a traversable wormhole, but \textit{only} in its Ellis III
reincarnation. We believe that this argument provides some understanding of
the \textquotedblleft other side":\ The mathematical operations of conformal
rescaling plus Wick rotation plus a trigonometric relation eases out the
naked singularity and converts the Brans-Dicke I solution into a traversable
wormhole.

The next task was to investigate into the extended solutions with new
parameters. Thus, in section VI, we studied the geodesic motion in the
extended geometry and obtained remarkable new results in the Einstein
minimally coupled theory: For nonzero values of constant $\theta $, the
spacetime acts like a radial sink. For $\theta =0$, the spacetime allows
nonradial motions ($U^{3}\neq 0$) but the Matos-N\'{u}\~{n}ez parameter $a$
can be entirely expressed in terms of the constants of motion. In Sec. VII,
we calculated the Sagnac effect in the extended spacetime and found that the
delay depends on $a$. If $a=0$, the delay is zero implying that $a$ could be
interpreted as a rotation parameter thereby supporting the conclusion of
Matos and N\'{u}\~{n}ez [8] from an altogether different viewpoint.

A very pertinent question is the stability of wormholes in the scalar-tensor
theory, especially in the Einstein minimally coupled theory under
consideration here. To our knowledge, the instability of black hole
solutions in the Einstein conformally coupled theory was shown first by
Bronnikov and Kireyev [23]. As to the wormhole solutions, it has been shown
by Bronnikov and Grinyok [17] that they are also unstable under spherical
perturbations $\delta \varphi $ of the scalar field satisfying $\varphi
_{;\mu }^{:\mu }=0$. Instabilities occur due to the occurrence of negative
energy levels in the effective potential. The authors suspected that such
instabilities could be generic features of scalar-tensor theories.
Instability of massive wormholes in the Einstein minimally coupled theory
has been shown by Armendariz-P\'{\i}con [17]. However, he also showed that
small mass wormholes in that theory could still be stable. It should be of
interest to study whether the extended solutions are stable under
perturbations. This remains a task for the future.

\textbf{Acknowledgments:}

The authors are deeply indebted to Guzel N. Kutdusova of BSPU for efficient
and useful assistance. We thank three anonymous referees for useful
criticisms and suggestions.

\textbf{References}

[1] M.S. Morris and K.S. Thorne, Am. J. Phys. \textbf{56}, 395 (1988); M.S.
Morris, K.S. Thorne and U. Yurtsever, Phys. Rev. Lett. \textbf{61}, 1446
(1988).

[2] A. Einstein and N. Rosen, Phys. Rev. \textbf{48}, 73 (1935).

[3] D. Hochberg, Phys. Lett. B \textbf{251}, 349 (1990).

[4] M. Visser, \textit{Lorentzian Wormholes- From Einstein to Hawking}
(A.I.P., New York, 1995).

[5] A.G. Agnese and M. La Camera, Phys. Rev. D \textbf{51}, 2011 (1995);
K.K. Nandi, A. Islam and J. Evans, Phys. Rev. D \textbf{55}, 2497 (1997).

[6] C. Barcel\'{o} and M. Visser, Class. Quant. Grav. \textbf{17}, 3843
(2000); B. McInnes, J. High Energy Phys. \textbf{12}, 053 (1002); G.
Klinkhammer, Phys. Rev. D \textbf{43}, 2512 (1991); L. Ford and T.A. Roman,
Phys. Rev. D \textbf{46}, 1328 (1992); \textit{ibid}, \textbf{48}, 776
(1993); L.A. Wu, H.J. Kimble, J.L. Hall and H. Wu, Phys. Rev. Lett. \textbf{%
57}, 2520 (1986).

[7] K.K. Nandi, B. Bhattacharjee, S.M.K. Alam and J. Evans, Phys. Rev. D
\textbf{57}, 823 (1998); L. Anchordoqui, S.P. Bergliaffa and D.F. Torres,
Phys. Rev. D \textbf{55}, 5226 (1997).

[8] T. Matos and D. N\'{u}\~{n}ez, Class. Quant. Grav. \textbf{23}, 4485
(2006) and references therein; T. Matos, Gen. Rel. Grav. \textbf{19}, 481
(1987).

[9] C.H. Brans, Phys. Rev. \textbf{125}, 2194 (1962).

[10] A. Bhadra and K.K. Nandi, Mod. Phys. Lett. A \textbf{16}, 2079 (2001).

[11] A. Bhadra and K. Sarkar, Gen. Rel. Grav. \textbf{37}, 2189 (2005).

[12] S. Kar, Class. Quant. Grav. \textbf{16}, 101 (1999)

[13] Y.M. Cho, Phys. Rev. Lett. \textbf{68}, 3133 (1992).

[14] G. Cl\'{e}ment, Phys. Lett. B \textbf{367}, 70 (1996); Grav. \& Cosmol.
\textbf{5}, 281 (1999).

[15] C. Martinez, C. Teitelboim and J. Zanelli, Phys. Rev. D \textbf{61},
104013 (2000).

[16] H.A. Buchdahl, Phys. Rev. \textbf{115}, 1325 (1959).

[17] C. Armend\'{a}riz-P\'{\i}con, Phys. Rev. D \textbf{65}, 104010 (2002).

[18] K.K. Nandi, Y.-Z. Zhang, Phys. Rev. D \textbf{70}, 044040 (2004); K.K.
Nandi, Y.-Z. Zhang and K.B. Vijaya Kumar, Phys. Rev. D \textbf{70}, 127503
(2004).

[19] H.G. Ellis, J. Math. Phys. \textbf{14}, 104 (1973), \textit{ibid},
\textbf{15}, 520E (1974).

[20] J. Cris\'{o}stomo and R. Olea, Phys. Rev. D \textbf{69}, 104023 (2004).

[21] R. Olea, Mod. Phys. Lett. A \textbf{20}, 2649 (2005).

[22] O. Bertolami, Phys. Lett. B \textbf{234}, 258 (1990).

[23] A. Tartaglia, Phys. Rev. D \textbf{58}, 064009 (1998); K.K. Nandi, P.M.
Alsing, J.C. Evans and T.B. Nayak, Phys. Rev. D \textbf{63}, 084027 (2001);
A. Bhadra, T.B. Nayak and K.K. Nandi, Phys. Lett. \textbf{A295}, 1 (2002).

[24] M. Visser and D. Hochberg, in Proceedings of the Haifa Workshop on
\textit{\textquotedblleft The internal structure of black holes \ and space
time singularities", }Jerusalem, Israel, June, 1997, [gr-qc/970001], p.20.

\bigskip

\textit{[16] The solutions have been rediscovered repeatedly after they were
first derived by I.Z. Fisher, Zh. Eksp. Teor. Fiz. \textbf{18}, 636 (1948).}

\textit{[17] The present authors have found the above earliest reference
from K.A. Bronnikov and S. Grinyok, Grav. \& Cosmol. \textbf{7}, 297 (2001).}

\textit{[18] H.A. Buchdahl}

\textit{[19] K.A. Bronnikov, Acta Phys. Polon. \textbf{B4}. 251 (1973). The
form of the solution given herein in \textquotedblleft harmonic" coordinates
(}$t,u,\theta ,\varphi $) \textit{can be easily transferred to the
Janis-Newman-Winnicour \textquotedblleft standard" form (}$t,\rho ,\theta
,\varphi $) \textit{which is equivalent to the Buchdahl \textquotedblleft
isotropic" form. See, respectively, Refs.[17] and A.Bhadra and K.K.Nandi,
Int.J.Mod.Phys. \textbf{A16}, 4543 (2001); Janis-Newman-Winnicour, PRL}

\textit{[20] K.A. Bronnikov and G.N. Shikin, Grav. \& Cosmol. \textbf{8},
107 (2002).}

\textit{[21] We thank an anonymous referee for pointing this out.}

\textit{[22] The factor }$\sqrt{2}$\textit{\ appears because we took }$%
\alpha =+1$\textit{\ instead of }$+2$\textit{. Other identifications of the
parameters with the original solution in [19] are: (}$m,a,n$\textit{) of
Ellis are our (}$M,m,m\sqrt{1+\beta ^{2}}$\textit{).}

\textit{[23] K.A. Bronnikov and Yu. N. Kireyev, Phys. Lett. A67, 95 (1978). }

\begin{center}
\textbf{Appendix}
\end{center}

A wormhole is defined to be traversable by a hypothetical traveler if it
satisfies some general constraints [1]. We shall demonstrate that the Ellis
III wormhole satisfies all of these. It can be made traversable even by a
human traveler under suitable choices of constants $m$ and $\beta $. Let us
put one branch, say, the $+ve$ branch of Eq.(29) into the metric (24) and
rewrite it in the standard MTY form [1] by defining a radial variable $\rho $
as%
\begin{equation}
(l^{2}+m^{2})\exp [2\beta \{\frac{\pi }{2}-\arctan (\frac{l}{m})\}]=\rho
^{2}.  \tag{A1}
\end{equation}%
(Note that $l\rightarrow \pm \infty $ implies $\rho \rightarrow \pm \infty $
and $l\rightarrow \pm 0$ implies $\rho \rightarrow \pm me^{\pi \beta }$.)
Then the metric (24) in the coordinates ($t,\rho ,\theta ,\psi $) becomes%
\begin{equation}
ds^{2}=-e^{2\Phi (\rho )}dt^{2}+\frac{d\rho ^{2}}{1-\frac{b(\rho )}{\rho }}%
+\rho ^{2}(d\theta ^{2}+\sin ^{2}\theta d\psi ^{2})  \tag{A2}
\end{equation}%
where the redshift function $\Phi $ is%
\begin{equation}
\Phi (\rho )=\beta \left[ \arctan \left\{ \frac{l(\rho )}{m}\right\} -\frac{%
\pi }{2}\right]  \tag{A3}
\end{equation}%
and the shape function $b$ is
\begin{equation}
b(\rho )=\rho \left[ 1-\frac{[l(\rho )-m\beta ]^{2}}{\rho ^{2}}\exp [2\beta
\{\frac{\pi }{2}-\arctan (\frac{l(\rho )}{m})\}]\right] .  \tag{A4}
\end{equation}%
General constraints on $b$ and $\Phi $ to produce a traversable wormhole are
satified by the functions in (A3) and (A4). It may be verified that: (1)
Throughout the spacetime, $1-\frac{b(\rho )}{\rho }\geq 0$ and $\frac{b(\rho
)}{\rho }\rightarrow 0$ as $\rho \rightarrow \pm \infty $ (2) Throat occurs
at the minimum of $\rho $ where $b(\rho )=\rho $. This minimum $\rho _{0}$
corresponds to $l_{0}=m\beta $ so that
\begin{equation}
\rho _{0}=m(1+\beta ^{2})^{\frac{1}{2}}\exp [\beta \{\frac{\pi }{2}-\arctan
\beta \}].  \tag{A5}
\end{equation}%
(3) The spacetime (A2) has no horizon, that is, $\Phi $ is everywhere
finite. (4) The coordinate time $t$ measures proper time in asymptotically
flat regions because $\Phi \rightarrow 0$ as $\rho \rightarrow \pm \infty $.
(5) The spacetime has no singularities, as discussed in Sec.IV.

Some more constraints are necessary if the trip is to be undertaken by a
human traveler:\ (a) Trip begins and ends at stations located on either side
of the throat where the gravity field should be weak. This demands that (i)
the geometry at stations must be nearly flat, or, $\frac{b(\rho )}{\rho }\ll
1$, (ii) the gravitational redshift of signals sent from stations to
infinity must be small, or, $\left\vert \Phi \right\vert \ll 1$ and (iii)
the acceleration of gravity at the stations must be less than one Earth
gravity $g_{\oplus }=980cm\sec ^{-2}$, or, $\left\vert c^{2}(1-\frac{b}{\rho
})^{\frac{1}{2}}\frac{d\Phi }{d\rho }\right\vert \lesssim g_{\oplus }$.
While the first two constraints (i) and (ii) are easily met in virtue of the
general constraints (1) and (4) respectively, (iii) gives
\begin{equation}
\left\vert \frac{m\beta }{l^{2}+m^{2}}e^{\Phi }\right\vert \lesssim .\frac{%
g_{\oplus }}{c^{2}}.  \tag{A6}
\end{equation}%
For fixed finite values of $m$ and $\beta $, $e^{\Phi }\rightarrow 1$ for
large $l$ and hence this constraint can be easily satisfied at the stations.
(b) The tidal forces suffered by the human traveler should be tolerable
which means that the magnitude of the differential of four acceletation $%
\left\vert \Delta \overrightarrow{a}\right\vert $ should be less than $%
g_{\oplus }$ in the orthonormal frame ($e_{\widehat{0}^{\prime }},e_{%
\widehat{1}^{\prime }},e_{\widehat{2}^{\prime }}e_{\widehat{3}^{\prime }}$)
of the traveler. This constraint translates, for a traveler of length (head
to foot) $\sim 2meters$, into the following bounds on the components of
curvature tensor computed in his/her frame:%
\begin{eqnarray}
\left\vert R_{_{\widehat{1}^{\prime }}{}_{\widehat{0}^{\prime }}{}_{\widehat{%
1}^{\prime }}{}_{\widehat{0}^{\prime }}}\right\vert  &=&\left\vert \left( 1-%
\frac{b}{\rho }\right) \left( -\frac{d^{2}\Phi }{d\rho ^{2}}+\frac{\rho
\frac{db}{d\rho }-b}{2\rho (\rho -b)}\frac{d\Phi }{d\rho }-\left( \frac{%
d\Phi }{d\rho }\right) ^{2}\right) \right\vert   \TCItag{A7} \\
&\lessapprox &\frac{g_{\oplus }}{c^{2}\times 2metr.}\simeq \frac{1}{%
(10^{10}cm)^{2}}.  \notag
\end{eqnarray}%
This bound is meant to constrain $\Phi (\rho )$ which is already well
behaved at the throat $\rho =\rho _{0}$ where its value is $\beta \left[
\arctan \beta -\frac{\pi }{2}\right] $. This value is $-1$ as $\beta
\rightarrow \infty $, and $0$ as $\beta \rightarrow 0$. In general, as $\rho
\rightarrow \pm \infty $, $\Phi $ vanishes. Thus the constraint (A7) is
easily satisfied. This result is expected since all the curvature tensor
components fall off with distance and vanish at infinity [19]. Let us look
at the lateral bounds given by%
\begin{eqnarray}
\left\vert R_{_{\widehat{2}^{\prime }}{}_{\widehat{0}^{\prime }}{}_{\widehat{%
2}^{\prime }}{}_{\widehat{0}^{\prime }}}\right\vert  &=&\left\vert R_{_{%
\widehat{3}^{\prime }}{}_{\widehat{0}^{\prime }}{}_{\widehat{3}^{\prime
}}{}_{\widehat{0}^{\prime }}}\right\vert =\left\vert \frac{\gamma ^{2}}{%
2\rho ^{2}}\left[ \left( \frac{v}{c}\right) ^{2}\left( \frac{db}{d\rho }-%
\frac{b}{\rho }\right) +2(\rho -b)\frac{d\Phi }{d\rho }\right] \right\vert
\TCItag{A8} \\
&\lessapprox &\frac{g_{\oplus }}{c^{2}\times 2metr.}\simeq \frac{1}{%
(10^{10}cm)^{2}}  \notag
\end{eqnarray}%
which constrain the speed $v$ with which the traveler crosses the wormhole, $%
\gamma =(1-v^{2}/c^{2})^{-\frac{1}{2}}$. For the metric (A2), the above
works out to
\begin{eqnarray}
\left\vert R_{_{\widehat{2}^{\prime }}{}_{\widehat{0}^{\prime }}{}_{\widehat{%
2}^{\prime }}{}_{\widehat{0}^{\prime }}}\right\vert  &=&\left\vert R_{_{%
\widehat{3}^{\prime }}{}_{\widehat{0}^{\prime }}{}_{\widehat{3}^{\prime
}}{}_{\widehat{0}^{\prime }}}\right\vert =\left\vert \frac{\gamma ^{2}}{%
2\rho ^{2}}\left[ \left( \frac{v}{c}\right) ^{2}\left( \frac{2m(m+l\beta )}{%
l^{2}+m^{2}}\right) +\left( \frac{2m\beta (l-m\beta )}{l^{2}+m^{2}}\right) %
\right] \right\vert   \TCItag{A9} \\
&\lessapprox &\frac{g_{\oplus }}{c^{2}\times 2metr.}\simeq \frac{1}{%
(10^{10}cm)^{2}}.  \notag
\end{eqnarray}%
The maximum tidal force should be experienced by the traveler at the throat
at $l=m\beta $ or equivalently at $\rho =\rho _{0}$. Hence the velocity $v$
at the throat is determined by the following constraint%
\begin{equation}
\frac{\gamma ^{2}}{\rho _{0}^{2}}\left[ \left( \frac{v}{c}\right) ^{2}\right]
\lesssim \frac{1}{(10^{10}cm)^{2}}.  \tag{A10}
\end{equation}%
It is possible to adjust $m$ and $\beta $ such that we have a throat radius $%
\rho _{0}\sim 10metr.$ (say). Assuming that $v\ll c$ or $\gamma \sim 1$, we
obtain $v\lesssim 30metr./s(\frac{\rho _{0}}{10metr.})$ which shows that the
speed $v$ across the hole could be made reasonably small. (c) The traveler
should feel less than $g_{\oplus }$ acceleration throughout the trip which
requires%
\begin{equation}
\left\vert e^{-\Phi }\left( 1-\frac{b}{\rho }\right) ^{\frac{1}{2}}\frac{d}{%
d\rho }\left( \gamma e^{\Phi }\right) \right\vert \lesssim \frac{g_{\oplus }%
}{c^{2}}.  \tag{A11}
\end{equation}%
With $\gamma \sim 1$, this works out to the same constraint as in (A6) which
is already satisfied. (d) The total proper time interval $\Delta \tau $
measured by the traveler and the coordinate time interval $\Delta t$
measured at the stations for the entire trip should be about a year (say),
that is,%
\begin{equation}
\Delta \tau =\int_{-L_{1}}^{L_{2}}\frac{dL}{v\gamma }\lesssim 1year
\tag{A12}
\end{equation}%
\begin{equation}
\Delta t=\int_{-L_{1}}^{L_{2}}\frac{dL}{ve^{\Phi }}\lesssim 1year  \tag{A13}
\end{equation}%
where $L=-L_{1}$ and $L=+L_{2}$ are the proper distances of the stations
measured from the throat. Since $\gamma \sim 1$, $\Delta \tau \simeq
(L_{2}+L_{1})/\overset{-}{v}$ where $\overset{-}{v}$ is the average velocity
not exceeding $30metr./s$ for a $10$ $metr.$ throat radius. This suggests
that the total separation between the stations should be of the order of $%
9.4\times 10^{5}Km.$

It is expected that the rotating wormhole should also be traversable, the
only physical difference is that the travelers would feel an additional
centrifugal force. However, the mathematical treatment of traversability
criteria in the rotating case is quite cumbersome. We leave it as a future
task.

Consider the rotating solution, metric (7), with $f$ generated from the $+ve$
branch of Ellis III solution (29) given by
\begin{equation*}
f(l;m,\beta ,a)=\frac{8m\beta \delta \exp [2\beta \{+\frac{\pi }{2}-\arctan (%
\frac{l}{m})\}]}{a^{2}+4\delta ^{2}\exp [4\beta \{+\frac{\pi }{2}-\arctan (%
\frac{l}{m})\}]}.
\end{equation*}%
Since the metric components in (7) do not depend on $t$, the vector $\frac{%
\partial }{\partial t}$ is a Killing vector of the spacetime but is not
everywhere orthogonal to the constant $t-$ hypersurfaces. Let us use a time
variable $T$ by%
\begin{equation*}
dT=dt+a\cos \theta d\psi .
\end{equation*}%
The vector $\frac{\partial }{\partial T}$ is not a Killing vector but is
everywhere orthogonal to the constant $T-$hypersurfaces. Hence the metric
(7) becomes%
\begin{equation*}
ds^{2}=-fdT^{2}+f^{-1}\left[ dl^{2}+(l^{2}+l_{0}^{2})(d\theta ^{2}+\sin
^{2}\theta d\psi ^{2})\right] .
\end{equation*}%
Under the redefinition $dT=dt^{\prime }+\frac{\sqrt{1-f}}{f}dl$, the metric
(A) becomes%
\begin{equation*}
ds^{2}=-dt^{\prime 2}+(dl-\sqrt{1-f}dt^{\prime
})^{2}+(l^{2}+l_{0}^{2})(d\theta ^{2}+\sin ^{2}\theta d\psi ^{2}).
\end{equation*}%
It was shown in Eq.(63) of Sec.VI that one solution of the geodesic
equations was radial motion ($U^{3}=0$) $\ $on the plane $\theta =const$.
Without loss of generality we can choose $\theta =\pi /2$ which implies that
$dT=dt$ on the equatorial plane. The metric on the constant $t^{\prime }$
surface for $\theta =\pi /2$ is then $d\sigma
^{2}=dl^{2}+(l^{2}+l_{0}^{2})d\psi ^{2}$. This surface is isometric to a
catenoid $\{(x,y,z)\mid (x^{2}+y^{2})^{\frac{1}{2}}=l_{0}\cosh (\frac{z}{%
l_{0}})\}$ in $E^{3}$, the radius of the central hole being $l_{0}$ [19].
The catenoid with a central hole is the sink referred to in Sec.VI. The
generators of the catenoid are the radial trajectories of a hypothetical
traveler.

\end{document}